# Wide-area Wireless Communication Challenges for the Internet of Things

Harpreet S. Dhillon[*], Howard Huang[†], Harish Viswanathan[†]

[*]Virginia Tech (Email: hdhillon@vt.edu)
[†]Bell Labs, Alcatel-Lucent (Email: {howard.huang,harish.viswanathan}@alcatel-lucent.com)

April 13, 2015

ABSTRACT

Aided by the ubiquitous wireless connectivity, declining communication costs, and the emergence of cloud platforms, the deployment of Internet of Things (IoT) devices and services is accelerating. Most major mobile network operators view machine-to-machine (M2M) communication networks for supporting IoT as a significant source of new revenue. In this paper, we motivate the need for wide-area M2M wireless networks, especially for short data packet communication to support a very large number of IoT devices. We first present a brief overview of current and emerging technologies for supporting wide area M2M, and then using communication theory principles, discuss the fundamental challenges and potential solutions for these networks, highlighting tradeoffs and strategies for random and scheduled access. We conclude with recommendations for how future 5G networks should be designed for efficient wide-area M2M communications.

## 1. INTRODUCTION

The Internet of Things (IoT) consists of a network of physical devices connected with remote computational capabilities. By combining physical sensing with data analysis to create meaningful information, IoT platforms enable solutions in the realms of smart cities, smart grids, smart homes, and connected vehicles. It has been touted as an economic engine for growth as it increases productivity, reduces cost, and improves lives. The growth in the number of IoT devices deployed is accelerating as the concept gathers broader industry momentum. General Electric has estimated the impact of the "Industrial Internet" on the world economy to be about $15 trillion [1], and analysts project that IoT related addressable revenue for mobile network operators worldwide could be $255 billion by 2020.

The three main components of an IoT-enabled application are the devices, the network, and the application servers. The devices sense a physical characteristic of the environment (e.g., temperature or presence of an object) and send the information through a communication network. The data is aggregated and processed by servers to provide meaningful information or an actionable output. This output could be sent back through the network to trigger a set of actuator devices (air conditioning switch or alarm). In some cases, the output could involve human mediation, but in other cases, the entire system could operate autonomously. The communication network is often known as an M2M network to distinguish it from networks that relay traffic generated or consumed by humans. While the network in general consists of wired and wireless devices, the trend is for devices to be wirelessly connected to the network edge to enable lower-cost installation, easier physical reconfiguration, and mobile applications.

IoT applications using wireless communications are highly varied and differ in their requirements. From a networking perspective, classical IoT applications can be categorized along two dimensions of range and mobility. Range refers to the geographic spread of the devices. It describes whether the devices are deployed in a small area, say within a couple of hundred feet of each other, or are dispersed over a wider area. Mobility refers to whether the devices move and if so, whether they need to communicate while on the move. Table 1 shows the five categories of applications spanning several orders of magnitude differences in range. For each category, it shows the basic device characteristics, services and suitable networks.

| Application | Range | Mobility | Device characteristics | Service characteristics | Suitable networks |
|---|---|---|---|---|---|
| • Connected car<br>• Fleet management<br>• Remote health monitoring | ~1000m | Yes | Rechargeable battery | Managed service, highly secure | • Cellular<br>• Satellite |
| • Smart metering<br>• Parking meter | ~1000m | No | Low rate, low power, low cost | Managed service | • Cellular<br>• Dedicated network |
| • Hospital asset tracking<br>• Warehouse logistics | ~100m | Yes | Low rate, low power, low cost | Enterprise-deployed | • WiFi<br>• RFID |
| • Industrial automation<br>• Home automation | ~10m | No | Low rate, low power, low cost | Subscription-free | • Zwave<br>• Zigbee<br>• Wifi<br>• Powerline |
| • Personal activity<br>• Local object tracking<br>• Point of sale | ~1m | No | Low rate, low power, low cost | Subscription-free | • Bluetooth<br>• NFC |

**Table 1: M2M application categories. We focus on applications in the top two rows which have a required range of about 1000m for wide-area coverage. Applications in other rows have more established ecosystems.**

For localized IoT applications, a short-range network is most appropriate, allowing the use of unlicensed spectrum and maximizing battery life while meeting the networking needs. For example, many smart home applications for environment control and monitoring would be well served using an 802.11-based network. Shorter-range applications can be enabled using Bluetooth or NFC. The smartphone can be used as hub to enable personal IoT applications such as health monitoring and local object tracking. Bluetooth is often used to connect to IoT devices, and an 802.11 or cellular connection provides network access.

For wide-area IoT applications such as the connected car or fleet tracking, a mobile broadband network is more suitable because devices move over a wide area. For applications such as metering where the devices are widespread but there is little need for mobility, a wide area network is required but does not have to support seamless mobility. Although the mobile network meets the requirements for this category of applications, a dedicated network that is designed for low data rates without complex mobility management procedures can be significantly cheaper, have greater reach into buildings, and provide a substantially longer battery life for devices.

For the remainder of the paper, we focus on wide-area wireless M2M communication at the physical and access layers. In Section 2 we describe the challenges in more detail. In Section 3, we briefly describe M2M solutions in cellular standards and dedicated M2M networks. In Section 4, we consider the fundamental design strategies from information theory and communication theory perspective. We conclude with some thoughts on how future 5G cellular networks could be designed to accommodate M2M communication more efficiently.

## 2. WIDE-AREA M2M COMMUNICATION CHALLENGES

In order to meet the demands of high data rate applications such as video streaming, conventional mobile broadband cellular networks are designed for high system capacity, measured in terms of data rate per unit area (bps/km$^2$). This capacity can be increased by using more spectrum, increasing the density of base stations, or increasing the spectral efficiency (bps/Hz) of each base station. A typical LTE base station using 20MHz bandwidth would serve a few dozen active handset devices, each operating at a rate of up to several Mbps as opposed to a base station in a M2M network that needs to serve very large number of devices at low rates. In contrast to low cost IoT devices, handsets are capable of performing sophisticated signal processing and the handset battery can be charged frequently, even on a daily basis if needed.

The challenges of designing a wide-area M2M communication network are different from those of a conventional broadband network because of the characteristics of IoT applications and the constraints imposed by low-cost and low-complexity IoT devices. These differences affect the assumptions and performance metrics of the system design, and potentially motivate novel designs at the PHY and MAC layer. We highlight some key attributes of M2M networks and describe their impact on the system design.

- **Small payloads**. In conventional broadband streaming or high data rate applications, it makes sense to invest in control overhead to establish bearers for scheduled transmission. In many IoT applications such as meter reading or actuation, the payload could be relatively small (~1000 bits), consisting of an encrypted device ID and a measurement or actuation command [2]. For small payloads, the control overhead for scheduled transmission may not be justified, and thus the traditional connection oriented approach of establishing radio bearers prior to data transmission will be inefficient for M2M [3]. Different IoT applications could have different latency and reliability requirements, which will impact the optimal design. For example, a meter reading for water consumption would have a longer latency requirement than a sensor for detecting a basement flood condition.
- **Large number of devices**. The number of IoT devices per cell could be significantly larger than the number of mobile devices per cell if multiple devices are associated with each person, car and building and if additional devices are deployed throughout the environment [4]. For a given set of radio resources, more devices require improved efficiency of both the control and data planes.

- **Bursty demand**. Certain IoT applications could exhibit highly bursty and correlated service requests. For example, a sudden severe storm could activate a large number of flood sensors, which would otherwise send only infrequent heartbeat signals.
- **Extended range.** Some sensor and actuation applications would require coverage in areas beyond a conventional cellular network, like in basements. The system could be designed to account for an extended link budget of, say, 20dB. Alternatively, if the extended coverage is not required, the improved link budget could be used to reduce the M2M infrastructure required to serve a given cellular coverage area and to reduce the device cost by reducing the maximum device transmit power.
- **Enhanced device energy efficiency.** In contrast to conventional cellular devices, which can be charged on a daily basis, devices for many IoT applications may not be amenable for frequent charging [3]. Improving the device energy efficiency would reduce the operational expense of recharging or replacing batteries. One could also seek alternate energy sources such as energy harvesting from vibrational sources or light. These alternatives are often intermittent or unreliable, requiring novel approaches for resource allocation.
- **Reduced device cost**. Compared to conventional cellular devices, IoT devices would have limited functionality and should cost less. The devices would probably not use multiple antennas, and lower-quality RF components could result in degraded link budgets. Reduced baseband processing may not allow for sophisticated decoding or encoding and could also limit encryption techniques. With reduced complexity RF components, the devices could be restricted in the number of bands they operate on, possibly limiting global roaming capabilities.

With these insights, we now briefly describe M2M solutions in cellular standards and dedicated M2M networks in the next section.

## 3. Wide-area M2M Technologies

Mobile communication systems standardized in 3GPP have primarily targeted feature phones, smart phones and tablets. Over the last three decades multiple generations (2G/3G/4G) of technologies have been standardized and deployed for voice and data communications. The capability for data communication has been exploited to use these technologies for M2M communications. Among the 3GPP technologies, GSM is most widely used for M2M. Although GSM supports only low data rates, it is sufficient for many M2M applications, e.g., see [5]. Since GSM has been operational for over two decades, device costs are substantially lower than that for 3G and 4G. Furthermore, GSM typically has better coverage than 3G and 4G in most parts of the world.

Recently 3GPP has been considering a number of alternatives for M2M that are based on introducing modifications to the GSM and LTE standards to meet the M2M requirements of low complexity, larger coverage, and lower device cost. The two major proposals in GERAN on cellular IOT are briefly described below [6].

- **GSM enhancements for M2M.** Motivated by the low cost of GSM devices, some companies are in favor of making GSM/GPRS the *de facto* M2M technology. This approach seeks to make GSM more efficient for M2M by increasing uplink capacity, extending coverage of downlink for both control and data channels by 20 dB, reducing power consumption of M2M devices compared with legacy GSM devices and reducing complexity of the devices compared to that of the legacy devices. In order to allow more devices to transmit at the same time in the same frequency in uplink, multiplexing based on overlaid code division multiple access technique is proposed where orthogonal codes are used to separate the users simultaneously transmitting in the same time slot. Coverage is enhanced for control, synchronization, system information, and data channels and is essentially achieved through repetition, with different repetition levels based on the coverage class the device belongs to. Other enhancements include definition of new control messages with smaller payload sizes and introduction of a new lower power class.

- **Novel narrow band air-interface.** Some proponents are making a case for a new narrow band air-interface that is compatible with GSM channelization of 200 KHz. In this clean slate proposal, a new system is defined that is optimized for IoT. Narrow bandwidth channels based on frequency division multiplexing are defined both on the uplink and downlink within the 200 KHz bandwidth allowing frequency reuse with only 200 KHz of spectrum. Uplink channel bandwidth is 3.75 KHz with 5 KHz channel spacing and channel bonding of 2 or 4 channels is allowed for higher data rates. On the downlink channel bandwidth of 15 KHz is proposed. FDMA access is preferred over CDMA since synchronization and closed loop power control can be avoided. Block repetition and symbol spreading CDMA are also used for extended coverage. The base station operates in RF full duplex mode in order to maximize network capacity while the devices operate in half duplex mode to reduce the RF cost. Devices could either support GMSK modulation or linear modulation such as BPSK, QPSK and 8-PSK.

3GPP standards have also incorporated a number of enhancements to the LTE for M2M since Release 11. These enhancements include creating a low priority access indicator for MTC devices facilitating access barring of only MTC devices under overload, introduction of power saving state for increasing battery life of devices that only communicate sporadically, signaling load minimization for static devices by minimizing frequency of mobility signaling such as tracking area updates, and introduction of the MTC trigger function to wake up devices [7]. Additional enhancements are being studied in Release 13 with the objective of enhancing coverage and reducing device cost compared to existing LTE networks. In the proposed approach, a 1.4 MHz narrow band version of LTE will be used by MTC devices in downlink and uplink within any system bandwidth [8]. In order to reduce cost, the devices will support a reduced set of transmission modes, for example, excluding MIMO modes and high data rates, and will have reduced maximum transmit power of 20 dBm.

Recently, a number of specialized networks optimized for wide-area M2M communications have been developed for dispersed, static/portable devices with low throughput requirements. Examples include technologies from companies such as Sigfox, Semtech, and On-Ramp. These technologies use narrow bandwidth channelization and target extended range. Because of the extended range, a metropolitan area can be covered using fewer base stations. In addition, the dedicated M2M devices are often lower power, operate at lower clock rates, and have lower cost compared to 3GPP devices. While there are many vendors offering such specialized networks for M2M, there is no single global standard. Dedicated M2M networks operate in unlicensed spectrum, which could make service quality requirements difficult to guarantee. Table 2 shows a comparison of different MTC technologies. The data is collected from [9-11].

| FEATURE | LTE Rel 13 | Combined Narrow Band (NB) and Spread Spectrum (SS) (Semtech) | Cooperative Ultra Narrow Band (Sigfox) | Narrow Band M2M Clean Slate (Huawei/Neul) |
|---|---|---|---|---|
| Bandwidth | 1.4 MHz | 400 Hz to 12.8 KHz NB and 200 KHz SS UL / 3.2 KHz to 12.8 KHz DL | 160 Hz UL / 600 Hz DL | 2 or 3.75 KHz UL / 15 KHz DL per channel |
| UL Data Rate | TBD | 122 bps – 7.8 Kbps | 160 bps / 600 bps | 200 bps to 45 Kbps |
| Range / MCL | 155.7 dB (24 dBm Tx Pwr) | 164 dB (20 dBm TX Pwr) | 164 dB (24 dBm Tx Pwr) | 162 dB (24 dBm Tx Pwr) |
| Broadcast/Multicast | Yes | Yes | No | No? |
| Duplex | Full/Half Duplex (FDD) | Full-Duplex | Full Duplex | Full-duplex |
| Synchronization | Yes | Yes | No | Yes |

Table 2: Comparison of MTC technologies.

Massive machine communication is expected to be one of the main requirements for 5G. Since 5G will not be constrained by any backward compatibility, it is valuable to step back and study the problem from a fundamental communication theory perspective, which is done next.

## 4. FUNDAMENTALS OF WIDE-AREA M2M COMMUNICATIONS

### 4.1 Problem Statement

We assume a single cell with a base station at the center and M2M devices uniformly distributed in the cell. Since these devices are not always transmitting, the instants when they have data to communicate to the base station can be visualized as an arrival process at the base station. We model this is a Poisson process with mean $\lambda$. For simplicity, we assume each transmitting device has the same payload of L bits that needs to be communicated to the base station in time T using the system bandwidth of W Hz. The main goal of this section is to determine the transmit power per device needed to support a given arrival rate $\lambda$ at the base station as a function of W, T and L under various transmission strategies. This can be equivalently visualized as the *massive access management problem*, in which the goal is to determine the maximum arrival rate that can be supported for a given power constraint.

### 4.2 Transmission Approaches

Transmission approaches can be categorized into two main classes depending upon whether the devices are transmitting over dedicated resources or over a *shared* random access channel (RACH). The former is termed *scheduled transmission* and requires that the base station already has information about the number of devices (say K) requesting to transmit in the current *resource slice* along with their channel gains. The later is unscheduled (or simply RACH) transmission in which the base station does not have any information about the transmitting devices. In this work, we assume that while the base station does not know the exact number of devices in RACH, it can estimate the *average arrival rate* $\lambda$ from previous time slots, and that this estimate is accurate. In the classical systems, RACH is typically used to initiate a connection with the base station by transmitting control information, such as the device identity. However, in M2M communications, since the payloads are small, it may be "optimal" to send these payloads along with control information on RACH itself. As a result, we first look at the RACH and scheduled transmissions separately and then consider a two-stage design in which the control information is first sent over RACH and the data is then sent through scheduled transmission.

**4.2.1 Random Access Channel (RACH)**

*Optimal RACH strategy*. Given K devices that have data to transmit in the current resource slice, each transmits with probability $\Theta$. Each device encodes its data with one of Q randomly chosen codebooks, each consisting of $2^{L/(WT)}$ codewords. The message is prepended with short preamble that identifies the codebook. Each device transmits over full bandwidth W Hz for the slice duration of T sec. The receiver then jointly decodes the largest set of users whose rate vector is in the corresponding capacity region, while treating other users as interference. Optimizing this strategy over $\Theta$ results in throughput optimal uncoordinated performance. For other considerations, such as retransmissions, and more formal details, please refer to [12] where this strategy was recently proposed by the authors.

*Suboptimal RACH strategies*. The optimal strategy discussed above is computationally intensive, which renders it unfit for practical implementations. Therefore, we consider a few suboptimal but easily implementable strategies. In particular, we consider slotted CDMA, FDMA, or a hybrid FDMA-TDMA where the bandwidth is partitioned into bins of width Q Hz and the time slot size is optimized for each arrival rate to minimize transmit power. While open loop power control can be employed to compensate for path loss and shadowing, the transmit power requires an additional fade margin to ensure reliable transmission in the presence of fading.

Figure 1 shows the peak (95-th percentile) power of the three suboptimal strategies for transmitting a device payload of L = 500 bits with a total W = 100 KHz bandwidth in time T=1 second and with a 0.1 probability of failure [3]. The optimal strategy is also shown for reference. Devices are dropped uniformly in a cell of radius 2km, and the pathloss exponent is 3.7. If peak power is not constrained, then FDMA supports higher arrival rates. However, the bandwidth per FDMA bin may be impractically narrow at very high arrival rates. If the bin width is constrained to Q=1000Hz, a few dB of additional power is required by F-TDMA for moderate arrival rates of $\lambda$ = 100 per second. The power penalty is more significant with a Q=10KHz bin, which is similar to what could be achieved with LTE's 15KHz bins. Overall, CDMA has better power performance for arrival rates below its pole capacity. The larger the bandwidth slice allocated for a CDMA channel, the lower the transmit power. However, allocating the entire bandwidth results in reduced flexibility to adapt the bandwidth allocated to M2M. We thus recommend dividing the available bandwidth into channels of smaller bandwidths (for example, 100KHz) and operating multiple CDMA channels, as many as needed based on the traffic conditions. Such a channelized

CDMA can be implemented within a multi-carrier system with a suitable waveform such as UFMC [13] to suppress inter-channel interference.

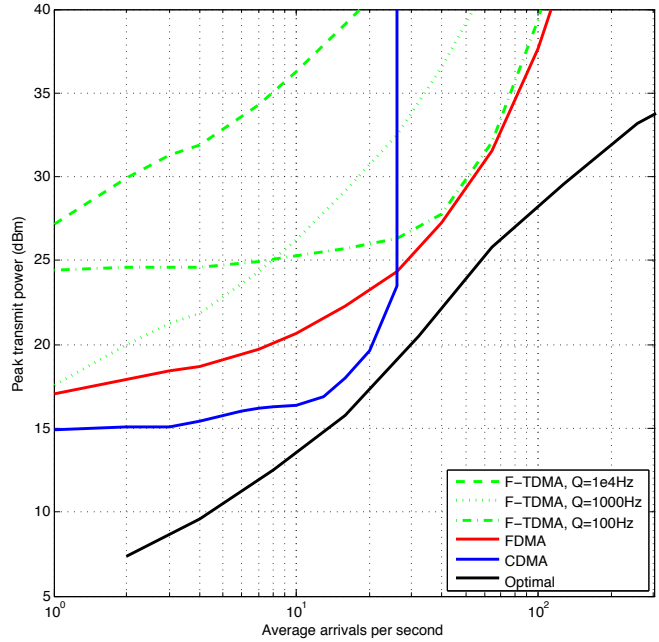

**Figure 1: Peak transmit power versus arrival rate performance for RACH transmission strategies, 100KHz bandwidth, 1 second latency, 500 bit payload per device.**

**4.2.2 Scheduled Transmission**

*Optimal scheduled strategy*. For given K devices, the optimal strategy is the one where all the devices transmit simultaneously over all the resources and the receiver performs weakest-last successive interference cancellation (SIC) [14]. Under this strategy, the receiver first decodes the device with the highest channel gain, assuming interference from the K-1 other devices. Using the decoded bits, the received signal for this device is reconstructed and subtracted from the received signal. Devices are decoded and cancelled successively in order from highest to lowest channel gains. The transmit power needed to communicate L bits in the given resource slice using this strategy was derived in [3].

*Suboptimal scheduled strategies*. The optimal strategy discussed above is sensitive to channel estimation errors. As was the case with the RACH transmission above, we consider more practical strategies using FDMA with either optimal or equal bandwidth allocation strategies [3]. Under optimal FDMA bandwidth allocation, W Hz bandwidth is allocated among the K devices to minimize the sum power. Under the equal bandwidth allocation, each device is allocated bandwidth W/K Hz. Using the same system assumptions as the RACH simulations (L = 500 bits, W = 100 KHz, T = 1 second, cell radius 2 Km, pathloss exponent 3.7), Figure 2 shows the peak (95-th percentile) power for the FDMA and the optimal SIC strategies. Comparing equal and optimal bandwidth allocation for FDMA, we note that equal allocation is near-optimal except at very high loading. While the optimal SIC strategy shows significant performance gains versus FDMA, it is important to note that its performance shown in Figure 2 assumes perfect channel estimates for all the devices which may not be realistic, especially at high arrival rates. In practice, the SIC performance gains are unlikely to be significant if we account for these impairments. This motivates the use of equal bandwidth FDMA as the preferred scheme for minimizing transmit power for the scheduled transmission. Note that FDMA could be implemented in practice as OFDMA in a wide band system.

While the scheduled strategies are able to achieve significantly higher packet arrival rates for a given transmit power constraint, it should be noted that control signaling is required to indicate a need and provide a grant for contention free resource allocation. We next compare the random access and scheduled strategies taking this overhead into account.

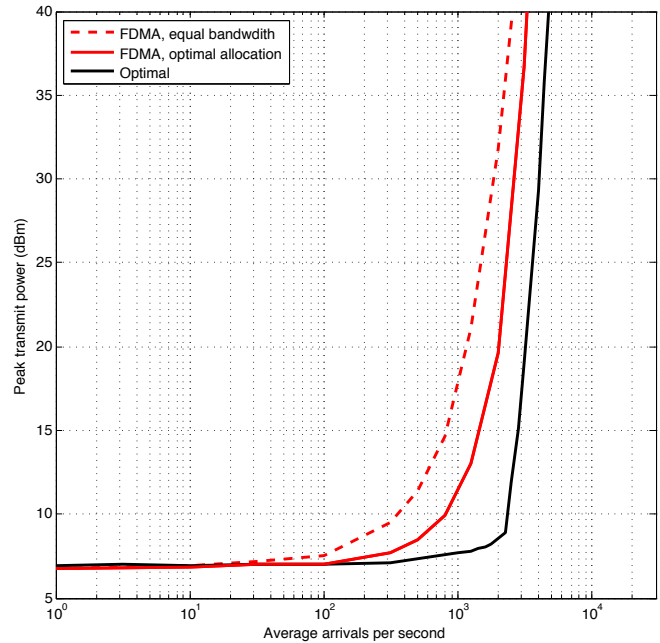

**Figure 2: Peak transmit power versus arrival rate performance for scheduled transmission strategies, 100KHz bandwidth, 1 second latency, 500 bit payload per device.**

### 4.2.3 One-stage vs. two-stage design

We focus on two system designs: (i) *one-stage* in which both data payload and control information are transmitted on uplink RACH, possibly resulting in a collission, and (ii) *two-stage* in which random access stage is used only for control information, and the data is communicated over contention-free scheduled resources in the second stage (RACH followed by scheduled transmission). The two protocols are presented in Figure 3 and their performance is compared in Figure 4. For the one-stage strategy, we consider either random access optimal or aloha FDMA transmission. For the two-stage strategy, we assume aloha FDMA for the first stage and scheduled FDMA for the second stage. For the purpose of calculating the control-signaling overhead on the downlink, we assume a downlink spectral efficiency of the system is 2.07 bps/Hz. For coordinated and uncoordinated transmission, we assume a maximum outage probability of 0.1, which includes the effect of retransmissions. The impact of finite block length codes are accounted for using the SNR gap values known in the literature [15]. *For smaller payload sizes, the supportable arrival rate for the one-stage strategies is higher than the arrival* *rate of the two-stage strategies because the two-stage overhead outweighs the relative inefficiency of the random access transmission. For larger payloads, the overhead becomes negligible, and the two-stage strategies are relatively more efficient.* For instance, for a payload size of 100 bits, the supportable arrival rate for optimal one-stage strategy is about one order of magnitude higher than that of the two-stage strategies. The crossover threshold between the one-stage aloha FDMA and two-stage strategies depends upon the level of overhead in two-stage strategy.

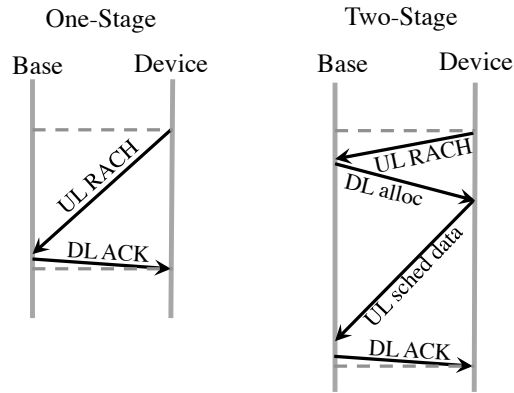

**Figure 3: Schematics of one-stage and two-stage protocols.**

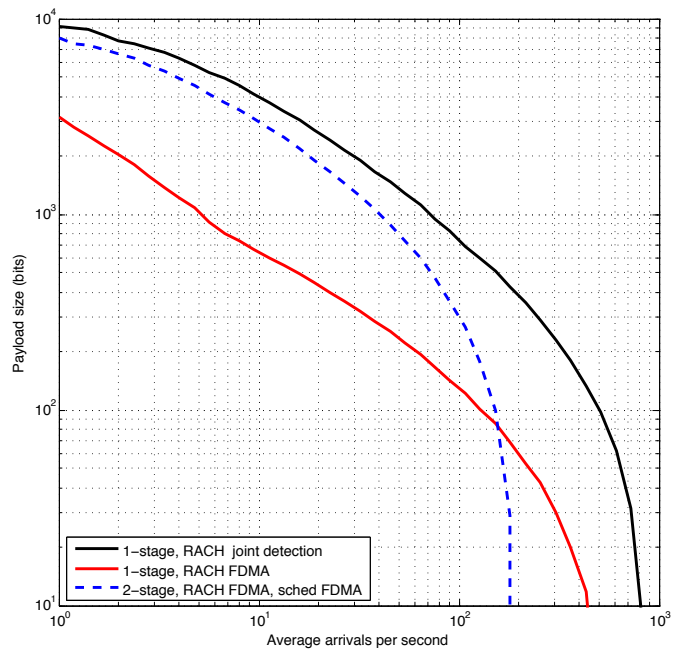

**Figure 4: Payload versus arrival rate performance for one-stage and two-stage transmission strategies, 10KHz bandwidth, 1 second latency.**

## 5. Conclusions

Current cellular networks are optimized for high-rate broadband access for sophisticated smartphone devices. To enable wide-area wireless communications for the future Internet of Things, the networks must also accommodate orders-of-magnitude more devices that communicate at orders-of-magnitude lower data rates. While evolving 2G and 4G cellular standards and dedicated networks for long-range M2M communication could address these needs in the near term, a future 5G cellular standard could present a unified solution that jointly optimizes the access network for both broadband and M2M communications. Our analysis suggests that for small payloads, a random access strategy with code multiplexing of transmissions within narrow bandwidth channels reduces transmit power and provides flexibility to allocate resources for such transmissions based on demand. For larger payloads, a scheduled transmission strategy carefully designed to minimize amount of control overhead is recommended. To provide additional functionality for an Internet of Mobile Things, the 5G standard could also be natively designed to enable ubiquitous localization and tracking for low-cost devices, in order to complement existing techniques such as GPS and RF fingerprinting.